\documentstyle[prl,aps,psfig]{revtex}

\begin{document}
\draft

\twocolumn[\hsize\textwidth\columnwidth\hsize\csname
@twocolumnfalse\endcsname

\widetext
\title{Theory of the Metal-Paramagnetic
Mott-Jahn-Teller Insulator Transition in A$_4$C$_{60}$  }
\author{Massimo Capone$^{(a,c)}$, 
Michele Fabrizio$^{(a,b,c)}$, Paolo Giannozzi$^{(d,e)}$,
 and Erio Tosatti$^{(a,b,c)}$}
\address{$^{(a)}$ International School for Advanced Studies (SISSA) ,
Via Beirut 2-4, I-34014, Trieste, Italy}
\address{$^{(b)}$ International Centre for Theoretical Physics (ICTP),
Trieste, Italy}
\address{$^{(c)}$ Istituto Nazionale Fisica della Materia (INFM)
Unit\`a Trieste-SISSA, Trieste, Italy}
\address{$^{(d)}$ Scuola Normale Superiore, Piazza dei Cavalieri 7, 
I-56126, Pisa, Italy}
\address{$^{(e)}$ Istituto Nazionale Fisica della Materia (INFM), 
Unit\`a Pisa-Scuola Normale Superiore, Pisa, Italy} 
\date{\today}
\maketitle

\begin{abstract}
We study the unconventional insulating state in A$_4$C$_{60}$ with a variety 
of approaches, including
density functional calculations and dynamical mean-field theory.
While the former predicts a metallic state, in disagreement 
with experiment, 
the latter yields a (paramagnetic) Mott-Jahn-Teller insulator. 
In that state, conduction between molecules is blocked by on-site 
Coulomb repulsion, magnetism is suppressed by intra-molecular 
Jahn-Teller effect, and important excitations (such as optical 
and spin gap) should be essentially intra-molecular. 
Experimental gaps of 0.5 eV 
and 0.1 eV respectively compare well with molecular ion values, 
in agreement with this picture.

\end{abstract}

]

\narrowtext
The alkali fullerides  A$_{n}$C$_{60}$, based on tightly bound
C$_{60}$ molecules, are relatively simple
molecular conductors where strong correlations, orbital degeneracy 
and Jahn-Teller effect intervene at the same time, leading to 
remarkable physical properties\cite{Gunn-rev}.
The three-fold degenerate $t_{1u}$ state of the isolated molecule 
leads to three bands hosting the valence electrons 
provided by the alkali metals. Hence all compounds A$_n$C$_{60}$ 
with $0<n<6$ should 
in principle be metallic if, as one expects, the bandwidth (of order   
0.5 eV), is larger than the relatively small crystal field splittings.
The $n=3$ fullerides 
are in fact generally metals, and become superconducting with T$_c$ as high 
as 40 K \cite{Ramirez}. Among the $n=4$ compounds, however, Na$_4$C$_{60}$, 
a stable fcc structure at high temperature\cite{Forro}, is the only one
which is also a paramagnetic metal, whereas K$_4$C$_{60}$ and
Rb$_4$C$_{60}$, stable bct structures down to low temperatures,
are instead paramagnetic narrow-gap insulators. Understanding in detail
this insulating state proves nontrivial, and constitutes
the main scope of this work. Known experimental parameters for
K$_4$C$_{60}$  
are a minimum band gap (probably indirect) between 0.05 and 0.2
eV\cite{musr,Ruani,Kerkoud},
and a direct, optical gap of 0.5-0.6 eV\cite{Knupfer}.
The insulating state is non magnetic with a spin gap
to the lowest triplet exciton of 0.1-0.14 eV\cite{Kerkoud,Mehring}.
Transition to a metal can be provoked by pressure\cite{Kerkoud},
which rationalizes why Na$_4$C$_{60}$ should be metallic, Na being 
the smaller cation.

One can invoke at least two possible scenarios for the insulating behavior of 
A$_4$C$_{60}$: a) a band insulator, due to a strong splitting of the 
$t_{1u}$ bands, arising for example from the bct distortion, or else from 
a collective static
Jahn-Teller (JT) distortion of the 
C$_{60}$ molecules, or alternatively b) a Mott-Jahn-Teller 
insulator\cite{Fabrizio}, where the hopping between adjacent molecules is 
first of all suppressed by a strong Coulomb repulsion, 
an intra-molecular JT effect subsequently optimizing the state
of the four localized electrons.  
We have studied first the molecular ion, and
then the possible 
origin of the insulating behavior of A$_4$C$_{60}$ by several theoretical 
approaches, tight-binding Hartree-Fock, density functional, and Dynamical 
Mean Field Theory (DMFT), the latter a powerful 
method to treat the Mott 
transition\cite{rev_dmft}. 

We start with the single C$_{60}^{4-}$ molecular ion. 
Within the
$t_{1u}$ orbital, assuming rotational (icosahedral) symmetry, 
a general interaction among electrons  
can be written as 
\begin{eqnarray}
H_{int}&& = 
\frac{U}{2} n^2 + \label{eq:Hint}\\
&&\frac{U_2}{6}\left[ 3\left(n_1-n_2\right)^2 + 
3\sum_{i<j}\Delta_{ij}^2 + \left(n_1+n_2-2n_3\right)^2\right],\nonumber
\end{eqnarray}
where  
$n_i = \sum_\sigma c_{i,\sigma}^\dagger c_{i,\sigma}^{\phantom{\dagger}}$ 
is the electron number on each orbital ($i=1,2,3$), $n=n_1+n_2+n_3$, and
$\Delta_{ij} =\sum_\sigma 
c_{i,\sigma}^\dagger c_{j,\sigma}^{\phantom{\dagger}} + H.c.$. At fixed $n$, 
the second term is responsible for the multiplet splitting. For $n=2,4$, 
the lowest energy state is the  $^3T_{1g}$ Hund's rule triplet, 
followed at energy 
$2U_2$ by a $^1H_g$ singlet, and at $5U_2$ by a $^1A_g$ singlet. 
The exchange coupling $U_2$ is expected not to be significantly screened 
by the $t_{1u}$ electrons, so that a reasonable estimate can be obtained 
by optical measurements on solid C$_{60}$, which give 
$U_2\simeq 0.05 \mbox{eV}$\cite{Sawa}.
Next, we consider the JT coupling to the eight $H_g$ vibrational modes.
In the adiabatic limit, one can use 
a single-mode approximation\cite{Manini2}
\begin{eqnarray}
H_{JT} &=& \frac{\hbar\omega_* g_*^2}{2} \left\{ \left(z^2+r^2\right)
+ \right.\nonumber \\
&& \left. z\left(n_1+n_2-2n_3\right) +r \sqrt{3} \left(n_1-n_2\right)\right\}.
\label{eq:HJT}
\end{eqnarray}
By using vibrational frequencies and couplings extracted from 
gas phase $C_{60}^{(-)}$ photoemission\cite{Gunn-ph}, we obtain 
$\hbar \omega_* = 0.117 \mbox{eV}$, and a dimensionless coupling 
$g_* = 1.204$, so that $E_{JT} = \hbar\omega_* g_*^2 = 0.169 \mbox{eV}$.
Instead of attempting an exact solution of $H_{int} + H_{JT}$, we
consider two opposite limits,
antiadiabatic and adiabatic. Although neither of them strictly
applies, for $\hbar \omega_*$ is comparable to all splittings and the
coupling is of medium strength, the former limit 
will yield the right symmetries,
while the latter will be quantitatively much more accurate.

In the antiadiabatic limit, the JT term (\ref{eq:HJT}) 
gives rise to a non-retarded 
electron-electron interaction, which can be absorbed into 
$U_2 \longmapsto U_2 - (3/4)E_{JT}$, with a change of sign of  $U_2$ 
from 0.05 to $ -0.076\mbox{eV}$.
The lowest energy state is now the $^1A_g$ singlet, followed at 
$0.23\mbox{eV}$ by the $^1H_g$ singlet, 
and at $0.38\mbox{eV}$ by the $^3T_{1g}$ triplet. The overall JT
energy gain in this limit is very large, $0.84\mbox{eV}$, about a factor three
larger than the bare adiabatic JT energy (see below).
This signals a true enhancement, due to the
gain in zero point energy corresponding to the frequency collapse
of the tangential vibron modes, first pointed out in 
Ref.\cite{Manini1}, as a possible mechanism for 
explaining the high 
critical temperature of A$_3$C$_{60}$ compounds.
 
In the adiabatic limit we diagonalize 
(\ref{eq:Hint}) plus (\ref{eq:HJT}) for $n=4$, 
minimizing successively the eigenvalues with respect to $z$ and $r$ 
treated as classical variables. We find 
a lowest energy singlet  
at a  classical distortion $z = -1.987$ and $r=0$, gaining 
$E_{JT} =0.293\mbox{eV}$. 
Adding the zero point energy gain  
$\hbar\omega_*$\cite{Manini1}, we obtain a total gain of  
$0.41\mbox{eV}$, the zero-point enhancement still sizable.
(A similar total gain of $0.42\mbox{eV}$ was obtained by 
uncorrelated eight-mode calculations\cite{Gunn-ph,Manini2}.)
The lowest triplet state has instead $z=1.0$ and $r=0$ and lies
above the ground state by $E_t = 0.108\mbox{eV}$ (spin gap). 
The lowest singlet, (with $z=1.0$ and $r=0$), is at
$E_s = 0.208 \mbox{eV}$ above the ground state.
Finally, the {\sl optical} gap, identified with the 
JT orbital splitting, is $\Delta = (3/2) z E_{JT} \simeq 0.504 \mbox{eV}$.

Both limits therefore predict that the C$^{4-}_{60}$ ion should be a singlet.
If hybridization between adjacent molecules were much 
smaller than all the molecular gaps involved, then
a lattice of C$_{60}^{4-}$ molecules would indeed be a non magnetic 
insulator, moreover with an optical gap $\Delta$ and a spin gap 
$E_t$ remarkably close to
the experimental ones. In that case, an electronic structure
calculation and total energy minimization for the A$_4$C$_{60}$
lattice should yield a narrow $t_{1u}$ band split by an insulating gap,
in turn supported by a static (uniform or staggered) 
collective JT distortion of all molecules (scenario (a)).

A tight-binding Hartree-Fock (HF) approximation yields precisely that.
The Hamiltonian which we use is
\begin{eqnarray}
H = \sum_{i,j, a,b, \sigma} t_{ij}^{ab}c^{\dagger}_{ia\sigma}c_{jb\sigma} 
+ H_{int} +H_{JT},
\label{hamiltonian}
\end{eqnarray}     
where the hopping amplitudes $t_{ij}^{ab}$ are evaluated along the lines
outlined in Ref.\cite{Gunn-bande}, and the vibronic terms are treated in the 
adiabatic approximation, as in Eq.(\ref{eq:HJT}).
We look for a HF state with a non-zero uniform average of 
$n_1 = n_2 \neq n_3$ (collective JT state)
and find a stable band insulator
with direct gap and spin gap of $\sim 1.48 \mbox{eV}$ 
and an indirect gap
of $\sim 1.03 \mbox{eV}$, all of which are 
much larger than the experimental values.  
The calculated HF bandstructure is shown in Fig. 1 (upper panel). 
This indicates that this method, known to overestimate
insulating tendencies, is unreliable
for this problem, and more realistic first principles
calculations are called for in this case.

We have carried out a series of such calculations for K$_4$C$_{60}$,
starting with bct configurations compatible
with X-ray data\cite{Stephens}, 
and using state-of the-art 
pseudopotential plane-wave density functional techniques
\cite{Giannozzi}, with E$_{cut}$ = 35 Ry (55 Ry for refinements), 
and careful $k$-point summations. 
Confirming previous results\cite{Erwin}, we found 
first of all that intermolecular
electron hopping is not small, yielding metallic
$t_{1u}$ bands of width $W\simeq 0.6\mbox{eV}$.
We searched next for a spontaneous collective JT distortion
by relaxing atomic positions based on Hellmann-Feynman forces. 
In order to check that the calculation could in principle yield
such a delicate JT distortion, and also reproduce the molecular limit,
we performed test calculations at an artificially enhanced
intermolecular spacing of $13.3$\AA. In that case a distortion appeared
spontaneosly, leading to a distribution of carbon distances
from the C$_{60}$ center between 3.511 and $3.553$\AA. This distortion
magnitude $\Delta R = 0.042$\AA, although small, is very close to that reported
for [PPN(+)]$_2$C$_{60}$(2-) salts\cite{PPN}, namely $0.043$\AA, and that
is very gratifyng since JT distortions of C$^{-2}_{60}$ 
and C$^{4-}_{60}$ should be essentially the same.
The single-particle gap was only about 0.1 eV, instead of the
expected 0.5 eV, a standard density functional shortcoming of no
consequence for this case.
When carried out for the true structure (bct,intermolecular spacing $9.97$\AA)
however, an extensive search for a JT distortion with one molecule/cell
yielded no result, and the system remained undistorted and metallic, 
as shown in Fig. 1 (lower panel).
 
Notably, X-rays fail to find a static JT distortion\cite{Stephens}, 
however within a resolution of precisely $0.04$\AA, which is
inconclusive. Selected trial calculations with two molecules/cell 
also failed to yield a doubling,
as would be caused, e.g., by charge-density-waves \cite{Erwin}  
or by a staggered collective JT state.  We temporarily
conclude that, since
accurate density functional calculations 
cannot account for its insulating behavior, K$_4$C$_{60}$ is
probably not a band insulator. We will make this statement
stronger based on additional reasoning at the end of this paper.

Electron correlations can be strong in these compounds. A realistic 
estimate\cite{Gunn-rev,Sawa}, of the intra-molecular 
Hubbard $U$ of Eq.(\ref{eq:Hint}) is $\simeq 1.0\mbox{--}1.6\mbox{eV}$,
which is larger than the full 
bandwidth. The failure of density
functional calculations does not appear surprising in
this light.
To get a description of the insulating state, and of the insulator-metal
transition, we 
resort to dynamical mean-field
theory (DMFT)\cite{rev_dmft}, which  is exact in the limit of 
infinite coordination lattices, and has proved quantitatively
successful in describing the Mott transition.
 
Within DMFT, the Hamiltonian (\ref{hamiltonian}) is mapped
onto a three-fold degenerate impurity Anderson model, 
where the parameters describing
the conduction electrons, as well as their coupling  to the impurity, 
are determined self-consistently\cite{rev_dmft}.
Band structure enters the calculations via the
density of states (DOS) in the self-consistency equations.
We approximate the realistic DOS
with a semicircular DOS of same bandwidth $W$, appropriate for a 
tight-binding model on an infinite-coordination Bethe lattice.
This approximation does not change the qualitative
behavior of the model and is not expected to significantly influence
the value of the gap. The Anderson model is solved by means of
exact diagonalization with a finite number ($n_s$) of conduction
electron degrees of freedom\cite{caffarel}, checking convergence
as a function of $n_s$ by finite-size scaling.
We should look for either orbital symmetry broken and 
unbroken phases, as the capability to describe
a true Mott insulator without any symmetry breaking is a 
unique feature of DMFT, which is not shared by HF or density functional-based
calculations, always
implying symmetry breaking at metal-insulator transitions.
For simplicity, we have restricted our search for 
insulating solutions either without orbital symmetry breaking or with 
uniform JT ordering. 
In order to locate the metal-insulator transition (MIT), 
we calculate the quasiparticle residue
$Z$ (plotted in Fig. 2 as a function of $U/W$), 
whose vanishing identifies the critical 
$U/W$ above which a paramagnetic insulator is stable, and the metallic
one is not. 
Again, we limit our analysis to limiting cases:
{\it{(i)}} No JT coupling; {\it{(ii)}} Antiadiabatic JT effect; 
{\it{(iii)}} Adiabatic JT effect. 
{\it{Case (i)}} -- Orbital degeneracy without JT coupling, is well
studied for $U_2 = 0$\cite{Gunn-rev,Cox}. 
In the presence of a finite $U_2/W = 0.08$ 
we find a MIT at $U/W  \simeq 1.414 $ for symmetric solutions, much 
reduced with respect to $U/W\simeq 1.98$ with $U_2=0$.  
{\it{Case (ii)}} -- The JT coupling renormalizes the dipolar integral, 
leading to an effective $U_2/W = -0.127$. The critical MIT $U/W$ 
is shifted to a much lower value $U/W  \simeq 0.707 $.
{\it{Case (iii)}} -- 
Solution of DMFT equations is more involved,
since the self-consistency must be required only after averaging
over the classical vibrational (five-dimensional) variable $\vec Q$ 
\cite{millis,ciuk}. 
In the broken-symmetry case the average simplifies, at $T=0$, 
as the probability distribution of $\vec Q$ becomes a single 
$\delta$-function. We obtain a critical $U/W \simeq 0.9\,- \,1.0$
(the uncertainty due to convergence difficulty).
The orbitally  symmetric insulating solution, which becomes stable 
at $U/W \simeq 1.237$, is of particular interest since it describes 
a molecular insulator where each molecule is distorted with
equal probability in all possible directions and independently
from any other molecule. This state has therefore 
a very large entropy which could be reduced by including 
quantum fluctuations, for instance in the form of tunnelling
between the equivalent local distortions, a way of describing a 
dynamical JT effect, not included at the adiabatic level.
It should be noted that our adiabatic limit does not now include the 
vibron zero-point energy gain. Because it is present in the symmetric case 
only, it could lower the true critical $U/W$ value of this phase, which could 
in reality prevail over the broken symmetry case. Moreover, temperature 
would also favor the symmetric state, where entropy is higher.

We can now compare critical $U/W$'s with the true ones.
As before, we expect the adiabatic values
to be quantitatively more accurate. The
calculated critical values are in all cases substantially smaller
than the actual $U/W$ value, i.e. 1.7--2.7.
We conclude that the insulator is best explained as a Mott-Jahn-Teller state,
where orbital degeneracy has becomes split,
giving rise to an essentially intra-molecular $C_{60}^{4-}$ Jahn-Teller
state whose calculated gaps (optical, spin) agree
very closely with experimental data. In the adiabatic limit
the orbital symmetry is broken, but that is likely 
to change when zero-point energy and temperature are included.

The close agreement between the adiabatic gaps of $C_{60}^{4-}$  and
experimental optical and spin gaps in A$_4$C$_{60}$ is perhaps the strongest 
piece of evidence in favor of a Mott state. Suppose
one could even find, by some other band calculation, such as GW\cite{Louie},
a stable static collective JT state as in scenario (a). By necessity,
the collective JT distortion magnitude would have to be substantially 
smaller than that of the isolated molecular
ion, since electrons leaving the molecule very frequently to hop
on other molecules weaken the on-site JT effect.  But, in that case,
it should not be possible to observe optical and spin gaps of exactly
the right molecular magnitude, as one does. They would be much smaller,
corresponding to the delocalization, or spillout, of the band Wannier
function.  In the Mott state, the 
electron spillout to neighboring molecules is reduced to order
$t/U$, which is very small. Hence this is the only state that
can explain why JT electronic gaps are essentially intra-molecular.
When the Mott insulator state is destroyed, for example by pressure-induced
increase of $W$ and decrease of $U$,
this intra-molecular physics is expected to disappear rather
suddenly (giving way to a density functional-like metal),
instead of gradually as in a band MIT  transition.

Our conclusion is that the bct A$_4$C$_{60}$ are
not conventional band insulators, and that both strong correlations
and JT effects are crucial for their understanding. That, incidentally, 
suggests that also the n=3 superconducting 
fullerides, whose bandwidth is quite similar, are most likeky 
close to a strongly correlated state \cite{Sawa}.
In that case the metal competes in our picture with 
a Mott-Jahn-Teller spin 1/2 antiferromagnetic insulator, 
with the implication
that strong correlations should probably not be ignored when discussing 
superconductivity.

We thank wholeheartedly G. Santoro for his constant support, 
and S. Ciuchi and
FULPROP partners for information and helpful discussions. 
We acknowledge sponsorship 
from INFM (PRA HTSC), and from the European
Union, Contract No. ERBFMRX-CT97-0155 (FULPROP).

\begin{figure}
\centerline{\psfig{bbllx=80pt,bblly=200pt,bburx=510pt,bbury=650pt,%
figure=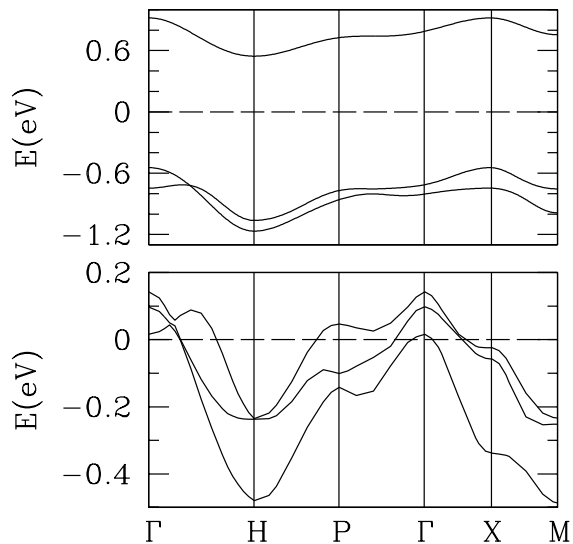,width=70mm,angle=0}}
\caption{
\label{fig_bande}
The $t_{1u}$ bands for K$_4$C$_{60}$ according to Hartree-Fock (upper panel) and
LDA (lower panel). The HF state
sustains a collective JT distortion and is wide gapped, 
the LDA state is undistorted, and metallic.
Both approximations disagree with experiment, indicating
a narrow gap semiconductor.}
\end{figure}

\begin{figure}
\centerline{\psfig{bbllx=80pt,bblly=200pt,bburx=510pt,bbury=575pt,%
figure=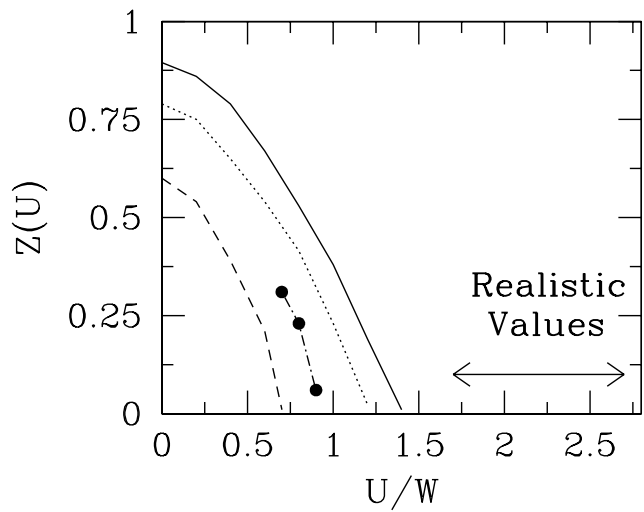,width=80mm,angle=0}}
\caption{
\label{fig_z}
Quasiparticle residue $Z$ as a function of $U/W$ for the 
symmetric solution of the purely electronic model 
(solid line), the symmetric (dotted line) and broken symmetry 
(dot-dashed line + heavy dots) solutions
in the adiabatic limit, 
and the symmetric solution
in the antiadiabatic limit (dashed line). 
The adiabatic broken symmetry value should be the best
numerical approximation. Realistic values of $U/W$ clearly indicate
a Mott-Jahn-Teller insulator.
}
\end{figure}

\end{document}